# Evidence for a preformed Cooper pair model in the pseudogap spectra of a $Ca_{10}(Pt_4As_8)(Fe_2As_2)_5$ single crystal with a nodal superconducting gap

Y. I. Seo[1], W. J. Choi[1], Shin-ichi Kimura[2] and Yong Seung Kwon[1,*]

[1]Department of Emerging Materials Science, DGIST, Daegu 711-873, Republic of Korea

[2]Graduate School of Frontier Biosciences and Department of Physics, Graduate School of Science, Osaka University, Suita 565-0871, Japan

For high-$T_c$ superconductors, clarifying the role and origin of the pseudogap is essential for understanding the pairing mechanism. Among the various models describing the pseudogap, the preformed Cooper pair model is a potential candidate. Therefore, we present experimental evidence for the preformed Cooper pair model by studying the pseudogap spectrum observed in the optical conductivity of a $Ca_{10}(Pt_4As_8)(Fe_2As_2)_5$ ($T_c = 34.6$ K) single crystal. We observed a clear pseudogap structure in the optical conductivity and observed its temperature dependence. In the superconducting (SC) state, one SC gap with a gap size of $\Delta = 26$ cm$^{-1}$, a scattering rate of $1/\tau = 360$ cm$^{-1}$ and a low-frequency extra Drude component were observed. Spectral weight analysis revealed that the SC gap and pseudogap are formed from the same Drude band. This means that the pseudogap is a gap structure observed as a result of a continuous temperature evolution of the SC gap observed below $T_c$. This provides clear experimental evidence for the preformed Cooper pair model.

* Corresponding author: E-mail address: yskwon@dgist.ac.kr (Yong Seung Kwon)

The BCS theory[1] announced in 1957 seemed to have revealed a universal law for material superconductivity by enabling understanding of the pairing mechanism for conventional superconductors. However, since it failed to explain the high-$T_c$ cuprate superconductors[2] discovered in 1986, we have encountered limitations in understanding the universal superconducting mechanism. Additionally, the phenomenon of a Mott insulator[3,4] or pseudogap[5-7] observed by the strong correlation in the normal state of the cuprate suggests that the pairing mechanism is caused by a more complicated process than the BCS theory[1]. Much effort has been made to understand this for cuprate superconductors. In particular, many attempts have been made to solve the mystery of the role and origin of the pseudogap seen in the temperature region higher than $T_c$. Pseudogap phenomena have also been found in heavy Fermion superconductors[8] and more recently in iron-based superconductors[9,10]. This implies that the pseudogap phenomenon is a common phenomenon in unconventional superconductors. Therefore, the correlation of the pseudogap in unconventional superconductors is expected to be closely related to the formation of superconducting electron pairs; the pseudogap mechanism must provide an important clue for understanding the universal pairing mechanism of unconventional superconductors.

There are many theoretical models[11-17] for explaining the origin of the pseudogap. Since the pseudogap is a partial gap at the Fermi surface that is observed above $T_c$ with an energy scale similar to the energy scale of the superconducting gap, among such models, the preformed Cooper pair model[18,19] is considered as a potential candidate. The preformed Cooper pair model describes a scenario where superconductivity is already present at the superconducting temperature $T_c$ or higher, but superconductivity does not appear due to its fluctuation. Therefore, many studies have been conducted to explain the relationship between superconductivity and the pseudogap phenomenon[5-7]. Recent studies have attempted to understand the pairing mechanism by explaining the pseudogap phenomenon experimentally observed in some iron-based superconductors by applying the theoretical preformed Cooper pair model[9,10]; however, such studies still lack clear experimental evidence to support the preformed Cooper pair scenario. Therefore, we have attempted to explain the pseudogap in the preformed Cooper pair scenario through experimental data obtained by using IR spectroscopy to observe and analyse the pseudogap spectrum.

In this study, we calculated the optical conductivity data by measuring the reflectivity of $Ca_{10}(Pt_4As_8)(Fe_2As_2)_5$ ($T_c$ = 34.6 K) single crystals at various temperatures and analysed the pseudogaps observed at temperatures of $T$ = 38, 70 and 100 K. We confirmed the temperature dependence of the Drude component and the definite gap spectrum in the low-frequency optical conductivity data for the temperature region where the pseudogap was observed. Through spectral weight analysis and analysis of gap spectrum for the superconducting and pseudogap regions, we demonstrate that the pseudogap is formed as a result of a continuous temperature evolution of the superconducting gap observed below $T_c$. As a result, we provide convincing experimental evidence that the pseudogap in the iron-based superconductor is a phenomenon that can be described by the preformed Cooper pair model.

**Results & Discussions**

Fig. 1 (a) shows the reflectivity spectrum data measured for $Ca_{10}(Pt_4As_8)(Fe_2As_2)_5$ (so-called Ca1048) single crystals from 8 to 300 K. The main panel is shown on a logarithmic scale from 10 to 10000 cm$^{-1}$, and the inset is shown on a linear scale from 0 to 200 cm$^{-1}$. Here, the data from 0 to 20 cm$^{-1}$ are not the measured values but are obtained by extrapolation, as described later. In the normal state ($T$ > 38 K), the reflectivity approaches unity as the frequency tends towards zero. This feature is more pronounced as the temperature is lowered, indicating that this compound shows the properties of a typical metal. When the superconducting state is entered at a temperature of 8 K, the reflectivity of the low-frequency region increases rapidly to a value of unity. However, flat reflectivity patterns such as those in Ba122[9] and Ca1038[10,20] iron-based superconductors are not observed. Generally, when an *s*-wave type superconducting gap is formed, it has a flat reflectivity characteristic in the low-frequency region. However, for the case of the Ca1048 compound, flatness in the reflectivity does not appear, indicating that the optical properties of the Ba122 and Ca1038 compounds are unlike those of the Ca1048 compound.

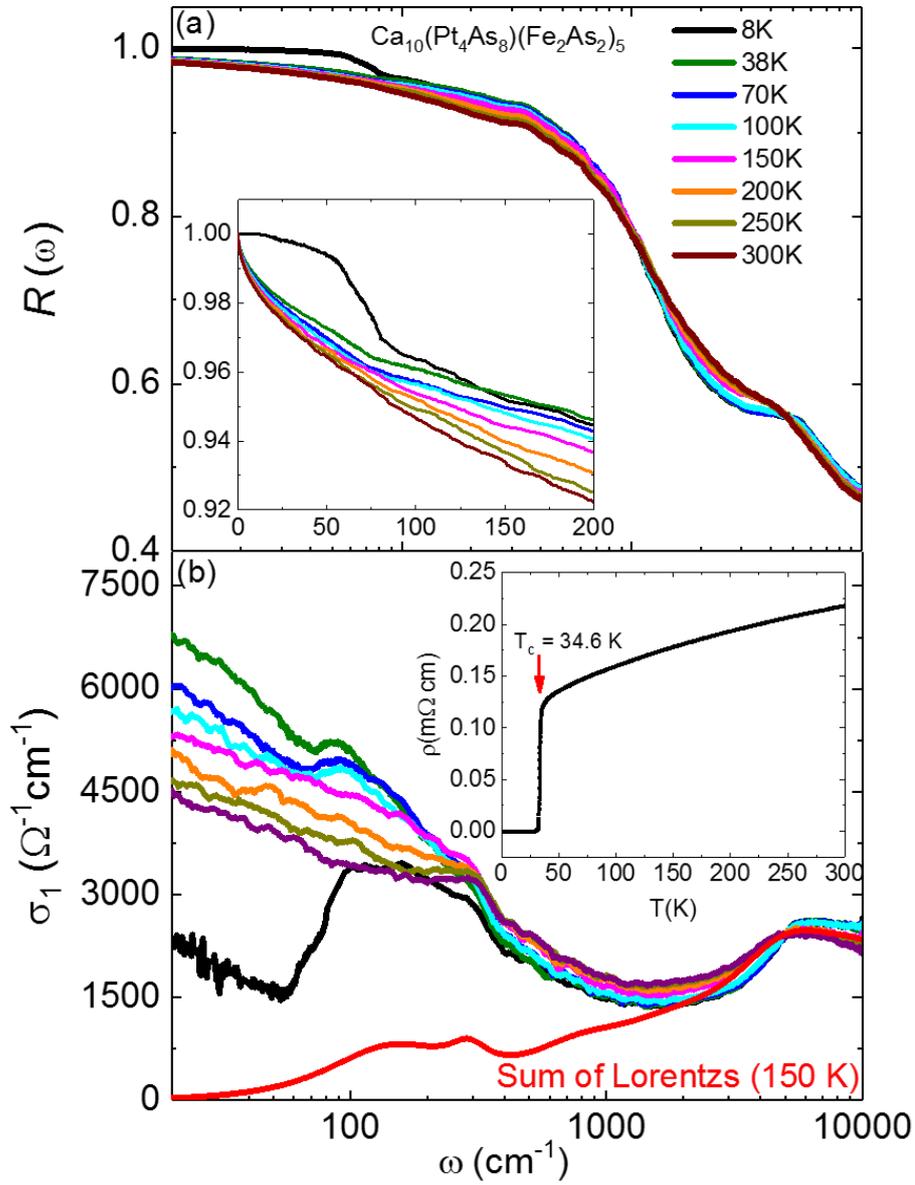

Fig. 1. (a) Reflectivity spectra $R(\omega)$ of $Ca_{10}(Pt_4As_8)(Fe_2As_2)_5$ single crystals at several temperatures. The inset graph shows an enlarged view of the region below a frequency of 200 cm$^{-1}$. (b) Frequency dependence of the real part of the optical conductivity $\sigma_1(\omega)$ for a $Ca_{10}(Pt_4As_8)(Fe_2As_2)_5$ single crystal at several temperatures. The red solid line represents the sum of the Lorentz oscillators at $T = 150$ K. The inset graph shows the temperature dependence of the DC electrical resistivity data for the $Ca_{10}(Pt_4As_8)(Fe_2As_2)_5$ single crystal.

For a more in-depth analysis, we have transformed the reflectivity $R(\omega)$ data into real-valued optical conductivity $\sigma_1(\omega)$ data using the Kramers-Kronig (KK) transformation. For the KK transformation, the Hagen-Rubens formula is used in the low-frequency region of the normal state, and (1-$A\omega^4$) is used in the superconducting state. On the other hand, in the region of high frequency above 12000 cm$^{-1}$, the reflectance is constant up to 40 eV and then extrapolated to the free-electron approximation $R(\omega) \propto \omega^{-4}$.

Fig. 1 (b) shows the optical conductivity obtained from the reflectivity on a logarithmic scale in the range of 0 to 10000 cm$^{-1}$. In the low-frequency region of the normal state, the optical conductivity data tend to become narrower as the temperature is lowered from higher temperatures. This is due to the development of a Drude component in the low-frequency range, which is consistent with the metal properties mentioned previously in the reflectivity data. Near 1000 cm$^{-1}$, the interband transition spectrum begins and reaches up to 10000 cm$^{-1}$. This interband transition spectrum shows a temperature dependence, as also observed in the optical conductivity data obtained for the La-doped Ca1038 sample reported previously[10]. The low-frequency optical conductivities at 38, 70 and 100 K show a hump at 0-100 cm$^{-1}$ along with the Drude spectrum. The simultaneous observation of a hump along with the Drude spectrum above $T_c$ is due to the partial gap formed in the Fermi surface. This is called the pseudogap; a pseudogap (PG) spectrum was also recently observed for a La-doped Ca1038 compound[10]. In the superconducting state at $T$ = 8 K, the optical conductivity is suddenly suppressed near the frequency of 100 cm$^{-1}$. This is because the opening of the superconducting gap has begun. However, the optical conductivity is not completely suppressed and has a residual Drude component. This phenomenon may be caused by the opening of superconducting gaps with nodes[21,22]. This will be discussed again later.

To analyse the optical conductivity represented by the contribution of the electron bands in the material, we used the standard Drude-Lorentz model as follows:

$$\sigma_1(\omega) = \frac{1}{4\pi}\left[\sum_j \frac{\omega_{p,j}^2}{\frac{1}{\tau_{D,j}} - i\omega} + \sum_k S_k \frac{\omega}{\frac{\omega}{\tau_{L,k}} + i(\omega_{0,k}^2 - \omega^2)}\right] \quad (1)$$

where $\omega_{p,j}$ and $1/\tau_{D,j}$ are the plasma frequency and scattering rate, respectively, for the $j$th free-carrier Drude band. $S_k$, $\omega_{0,k}$ and $1/\tau_{L,k}$ are the spectral weight, resonance frequency and scattering rate, respectively, of the *k*th Lorentz oscillator. As seen from the ARPES results[23,24], the Ca1048 compound shows a multiband structure similar to those of other iron-based superconductors[10,25], indicating that at least two Drude components associated with intraband transitions are needed in the optical conductivity fitting. Thus, we used two Drude components here. On the other hand, the optical conductivity caused by interband transitions was well fitted by six Lorentz components, similar to the fitting results for the La-doped Ca1038 compound[10]. The sum of the Lorentz components at 150 K, the temperature at which the pseudogap is not observed, is shown by the red solid line in Fig. 1 (b).

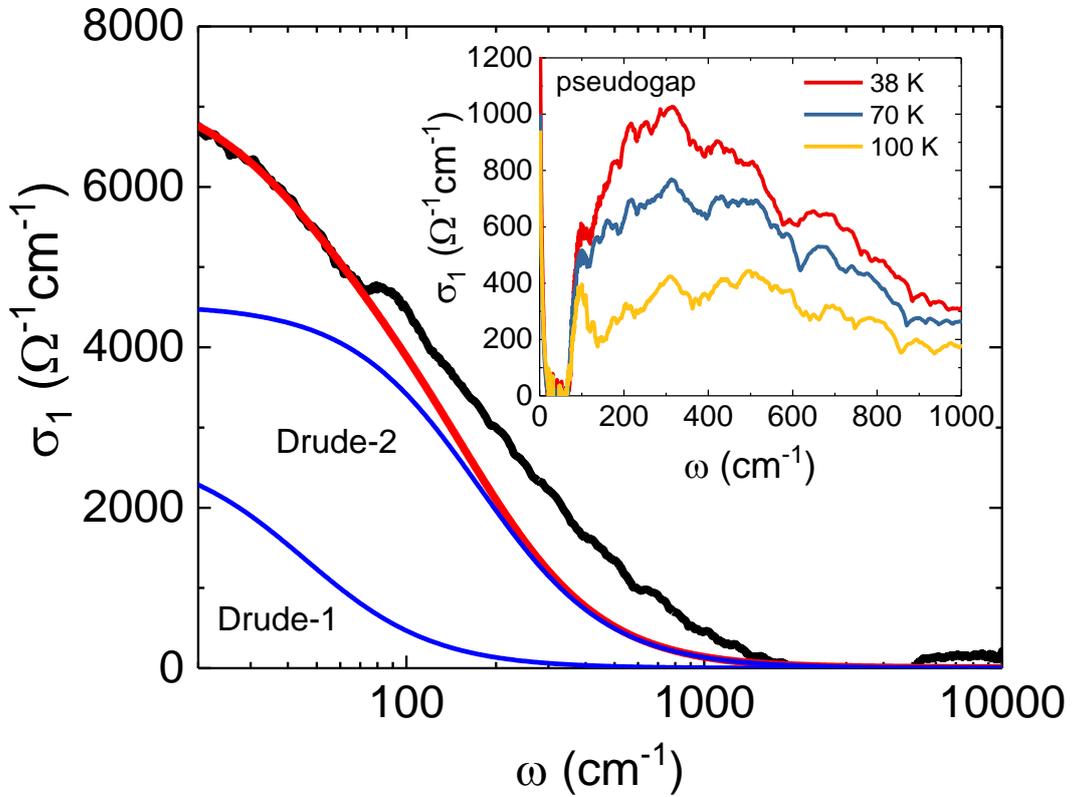

Fig. 2. The solid black line shows the optical conductivity with the subtraction of the sum of the Lorentz oscillators determined at $T = 150$ K from the optical conductivity at $T = 38$ K. The blue solid lines represent each Drude band, and the red solid line represents the sum of the Drude oscillators. The inset graph shows the temperature dependence of the pseudogap spectrum obtained by subtracting the red solid line from the black solid line.

To analyse only the PG spectrum in the optical conductivity data, we subtracted the high-frequency Lorentz sum obtained by the previous Drude-Lorentz model fitting from the data at $T = 38$, 70 and 100 K, where a PG was observed. The optical conductivity spectra due to a PG are shown as the solid black lines in Fig. 2. The spectra seem to be the sum of the spectrum of the Drude component and the spectrum of the gap type. To examine this in detail, the optical conductivities below 80 cm$^{-1}$ at 38, 70 and 100 K were fitted using two Drude functions, as shown in Fig. 2. The Drude sums (red solid line) agree well with the low-frequency experimental results due to the PG spectra. By removing these Drude sums from the PG spectra, we obtained gap spectra, which are plotted in the inset of Fig. 2. It was confirmed that this is a clear gap spectrum. In addition, a temperature dependence for the gap spectrum was clearly observed.

To observe the temperature evolution of the optical conductivity, the spectral weight was calculated as follows:

$$SW(T;\omega_c) = \int_{0^+}^{\omega_c} \sigma_1(\omega;T)d\omega = \frac{\pi^2}{Z_0}(\omega_p)^2 \qquad (2)$$

where $\omega_c$ is the cut-off frequency and $Z_0$ and $\omega_p$ are the vacuum impedance and plasma frequency, respectively. Fig. 3 (a) shows the temperature dependence of the spectral weight for each Drude band spectrum in the normal state and the PG spectrum in the pseudogap state. The spectral weight of the Drude-1 band drawn in blue squares is constant regardless of temperature in both the normal and PG states. The spectral weights of the Drude-2 band in the normal state, plotted as red solid squares, have a constant value regardless of the temperature in the temperature range of 300 to 150 K, whereas the spectral weights in the PG state, plotted as red open squares, decrease as the temperature is lowered from 100 K. The yellow open triangles in the figure show the temperature dependence of the spectral weight for the gap spectrum in the PG spectrum; it increases as the temperature decreases. The black open squares represent the total weight of the Drude-2 band and gap spectra in the PG state, which is approximately equal to the spectral weight of the Drude-2 band in the normal state above 150 K. Satisfaction of this sum rule indicates that the PG observed in the optical conductivity data is developed in the Drude-2 band and that no spectral weight is transferred to $\omega = 0$, unlike in the case of the superconducting Cooper pair. This suggests that the PG feature in this compound follows the preformed Cooper pair model. In the preformed Cooper pair model, there is superconducting (SC) correlation above $T_c$, but due to fluctuation in the SC correlation, the Cooper pairs are destroyed, which leads to the formation of a Drude-free band.

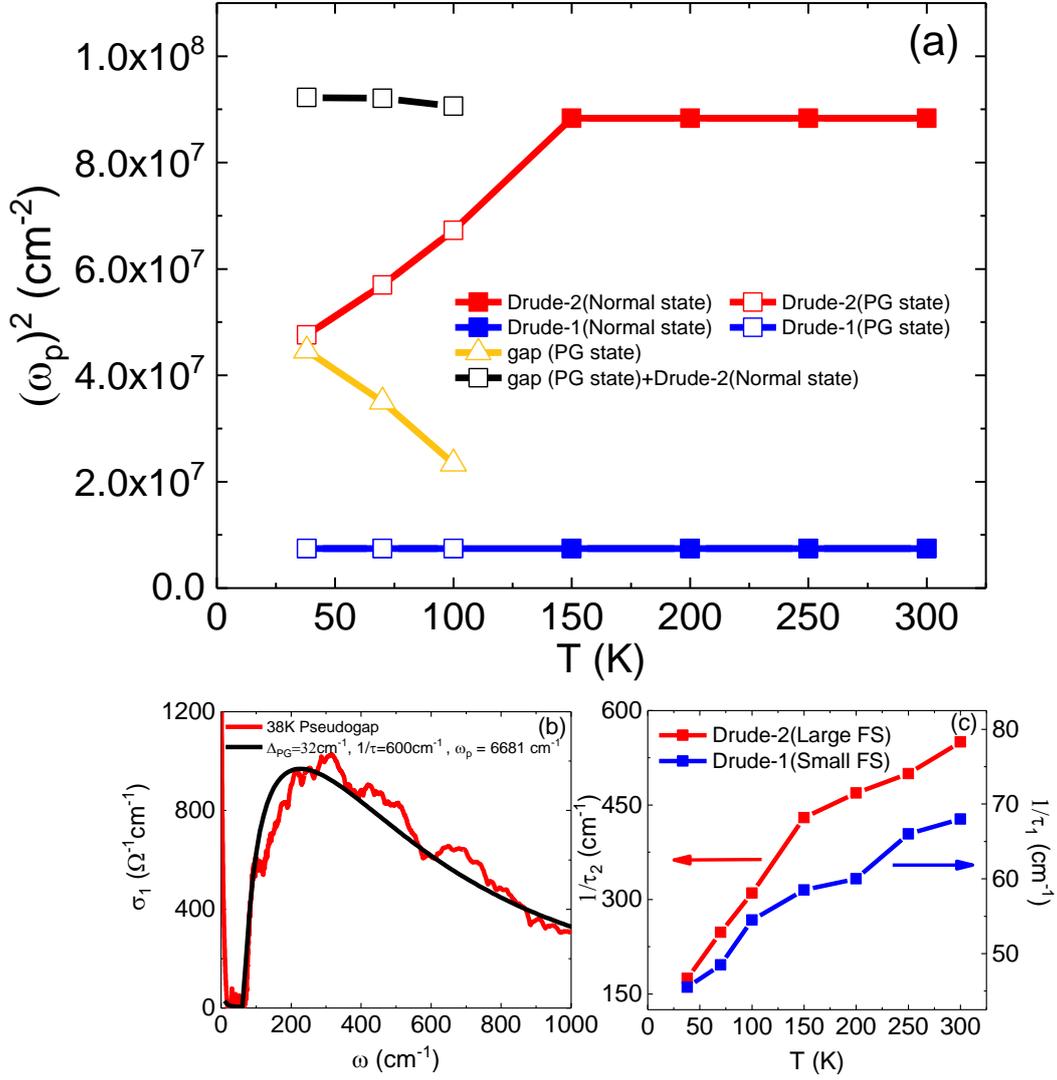

Fig. 3. (a) The graph shows the spectral weights of the pseudogap and Drude bands as functions of temperature. (b) The gap-type spectrum in the pseudogap spectrum at $T$ = 38 K fitted with the s-wave SC gap model[26]. The fitting parameters are $\Delta$ = 32 cm$^{-1}$, $1/\tau$ = 600 cm$^{-1}$ and $\omega_p$ = 6681 cm$^{-1}$. (c) The graph shows the scattering rate for each Drude band as a function of temperature.

Fig. 3 (b) shows the result of fitting the gap-type spectrum to the *s*-wave SC gap model[26] in the PG spectrum at 38 K obtained above. The fitting results are in good agreement with the experimental data, with fitting parameters of $\Delta$ = 32 cm$^{-1}$, $1/\tau$ = 600 cm$^{-1}$ and $\omega_p$ = 6681 cm$^{-1}$. Fig. 3 (c) shows the temperature dependence of the scattering rate for each Drude band. In both bands, the scattering rate decreases as the temperature decreases, with the Drude-2 band showing a larger scattering rate compared with the Drude-1 band. Fig. 4 shows the result of fitting the SC gap optical conductivity spectrum obtained by removing the interband transition spectrum for the superconducting state at $T = 8$ K using the *s*-wave Mattis-Bardeen model[26]. In this fitting, we used one gap (blue dotted line) and an additional Drude function (green dotted line). Optimally La-doped Ca1038 compounds reported previously[10] were described using two superconducting gaps. The parameters obtained from our fitting results are as follows: an SC gap size of $\Delta = 26$ cm$^{-1}$, a scattering rate of $1/\tau = 360$ cm$^{-1}$ and a normal-state Drude plasma frequency of $\omega_p = 8559$ cm$^{-1}$. The inset of Fig. 4 shows the temperature dependence of the SC gap and the spectral weights of the two Drude bands in the normal state. The pink inverted triangle and the green triangle represent the spectral weights of the SC gap and the extra Drude band, respectively, and the sum of the two spectral weights is represented by a black square. Since the total weight value is nearly equal to that of the Drude-2 band (red squares), it is clear that both the SC gap and the extra Drude component will occur in the Drude-2 band. Additionally, in a comparison of the pseudogap shown in Fig. 3 (b) with the SC gap shown just above, the gap spectra appear in almost similar spectral regions. As a result of the similarity of the energy scales for these two gaps and the satisfaction of the sum rule for the spectral weights, it is considered that the pseudogap occurs as a result of a continuous temperature evolution of the SC gap formed in a larger Fermi surface. This continuous change in temperature makes the preformed Cooper pair model scenario clearer.

As described above, in the superconducting state, optical conductivity is not completely suppressed but retains extra Drude components. This is also true for Co- or K-doped Ba122 compounds[21,22]. Optical spectroscopy studies of these compounds have shown that anisotropy of the SC gap may occur due to Fermi surface reconstruction caused by the spin density wave (SDW), which creates a node in the SC gap to show the residual Drude component. Based on these results, the residual Drude component should not be observed because the SDW is not present in the Ca1048 compound. However, since the residual Drude component is observed in the experimental results, the Ca1048 compound has an SC gap node. It is not possible to exclude correlation, such as short-range order SDW, as a cause for the existence of the SC gap node, as it is difficult to detect. Therefore, further studies are needed to clarify the apparent cause of the SC gap node in Ca1048 compounds.

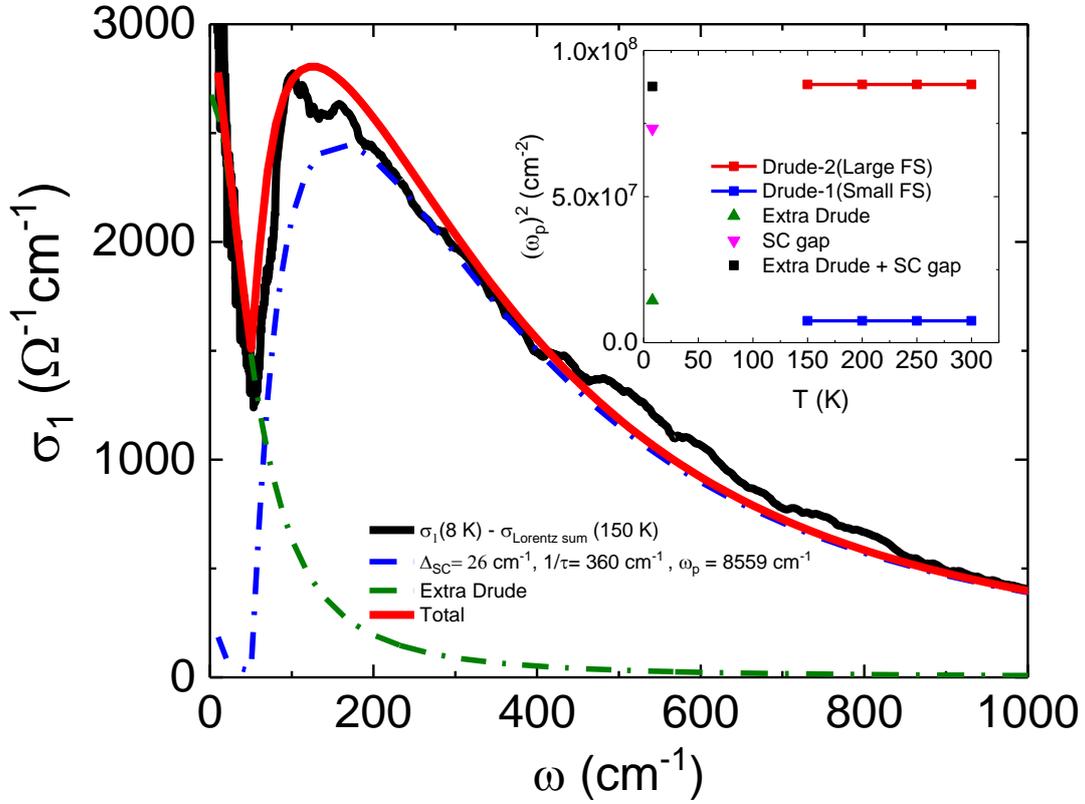

Fig. 4. The SC gap in the superconducting state at $T$ = 8 K was fitted with the $s$-wave Mattis-Bardeen model[26]. The blue dotted line represents the SC gap calculated with $\Delta$ = 26 cm$^{-1}$, $1/\tau$ = 360 cm$^{-1}$ and $\omega_p$ = 8559 cm$^{-1}$. The fitted extra Drude oscillator in the low-frequency range is represented by the green dotted line. The total sum of the data, including the SC gap and the extra Drude oscillator, is represented by the red solid line. The inset graph shows the spectral weights of the SC gap and Drude bands as functions of temperature.

From the data described above, we conclude that the PG and SC gaps are formed in the Drude-2 band with a large Fermi surface. In this compound, we need to consider whether the SC gap in the Drude-1 band is absent or not when observed via optical spectroscopy. As shown in Fig. 3 (c), the temperature-dependent scattering rate is much lower in the Drude-1 band than in the Drude-2 band, which suggests that the SC gap in the Drude-1 band is close to the clean limit[27]. According to the $s_\pm$-wave gap theory[28], a small SC gap is formed in the Drude band with a large Fermi surface, whereas a large SC gap is formed in the Drude band with a small Fermi surface. It is expected that the size of the superconducting gap in our samples will be larger in the Drude band-1 than in the Drude-2 band because this phenomenon is known to be in accordance with the $s_\pm$-wave gap theory for iron-based superconductors. Since the SC gap in the Drude-1 band is close to the clean limit, it is considered that the SC gap in the Drude-1 band will not be observed in optical conductivity measurements. The PG of the Drude-1 band is also not observed because of this clean limit condition.

In recent STM studies on Ca1048 compounds[29], the SC gap size was observed to be approximately 35 cm$^{-1}$. This result is larger than the size of the SC gap of the Drude-2 band mentioned above, so it is judged to be the SC gap formed from the Drude-1 band. This is because STM can observe the gap regardless of the presence or absence of the clean limit condition for the gap.

In summary, we measured the reflectivity spectra of $Ca_{10}(Pt_4As_8)(Fe_2As_2)_5$ single crystals and obtained the optical conductivity spectra. A pseudogap anomaly in these optical conductivity spectra was observed at $T$ = 38, 70 and 100 K. The pseudogap anomaly consists of a Drude response and a gap-type absorption. The spectral weights of the absorption spectrum for the gap and Drude bands in the PG gap spectrum were in agreement with those of the Drude-2 band in the normal state and satisfied the sum rule. This means that the Drude-2 band seen in the normal state exhibits a pseudogap in the temperature range below 100 K, which is normal. In the superconducting state at $T$ = 8 K, the optical conductivity can be explained by one SC gap by applying the $s$-wave Mattis-Bardeen model[26], and the increase in the optical conductivity in the low-frequency region caused by the SC gap node is explained by an extra Drude term. We found that the SC gap and PG were formed in the Drude-2 band because the sum rule for the spectral weights of the SC gap, the PG and the extra Drude component was satisfied. Therefore, we conclude that the PG observed above $T_c$ is formed as a result of a continuous temperature evolution of the SC gap from the evidence that the SC gap and PG are formed in the same Drude band and that the two gap spectra are observed in a similar frequency region. These results represent clear experimental evidence that the PG observed in iron-based superconductors is the Cooper pair breaking spectrum as described by the preformed Cooper pair model.

**Methods**

A single crystal of $Ca_{10}(Pt_4As_8)(Fe_2As_2)_5$ was grown using the Bridgman method with a sealed molybdenum (Mo) crucible and boron nitride (BN). First, the FeAs precursor was synthesized in evacuated quartz ampoules at 1050°C. Second, the FeAs precursor and the Ca and Pt elements were placed into a BN crucible; then, the BN crucible was placed into a Mo crucible, and a Mo lid was welded onto the crucible using an arc welder in a high-purity Ar-gas atmosphere. Finally, the entire assembly was slowly heated up to 1500°C in a vacuum furnace consisting of a tungsten meshed heater with a temperature stability of 0.1°C and kept at this temperature for 72 h; afterwards, the assembly was moved slowly at a rate of 1.8 mm/h in a downward direction for approximately 85 h, and then, finally, slowly cooled down to room temperature. As a result of this process, we obtained high-quality single crystals with a typical size of 2 x 2 x 0.5 mm$^3$.

The optical reflectivity spectra $R(\omega)$ of the $Ca_{10}(Pt_4As_8)(Fe_2As_2)_5$ single crystals were measured over frequency regions of 70 - 12000 cm$^{-1}$ and 20 - 150 cm$^{-1}$ using a Michelson-type and a Martin-Puplett-type rapid-scan Fourier spectrometer. The absolute $R(\omega)$ was determined by an *in situ* Au overfilling method[9]. In this method, the sample position was precisely sought with a feedback method using a He-Ne Laser and a Si-diode detector, reducing the reflectivity error by 0.3%. The real part of the optical conductivity $\sigma_1(\omega)$ and the dielectric constant $\varepsilon_1(\omega)$ were derived from the $R(\omega)$ spectra through the K-K transformation. In the K-K transformation, the reflectivity was extrapolated with a Hagen-Rubens function below 20 cm$^{-1}$, with a constant reflectivity from ~1.5 eV (= 12000 cm$^{-1}$) to 40 eV, and then with a free-electron approximation $R(\omega) \propto \omega^{-4}$.

**Acknowledgements**

Y.S.K. was supported by an NRF grant funded by the Ministry of Science, ICT and Future Planning (2016R1A2B4012672 and 2012K1A4A3053565) and by the International Joint Research Promotion Program of Osaka University (2016-2017).

**Author contributions statement**

Y.I.S. and W.J.C. performed the sample growth and the reflectivity experiment. S.K. assisted with the data analysis. Y.S.K. and Y.I.S. wrote the manuscript. All authors discussed the results and reviewed the manuscript.

**Additional information**

**Competing interests:** The authors declare that they have no competing interests.